\documentclass{article}
\usepackage{spconf,amsmath,graphicx,url,booktabs,amssymb,multicol,multirow,xcolor,pifont}
\usepackage[inline]{enumitem}
\usepackage{tablefootnote}
\usepackage{hyperref}
\usepackage{cleveref}

\newcommand{\cmark}{\ding{51}}%
\newcommand{\xmark}{\ding{55}}%

\definecolor{xiaomi_gray}{HTML}{A9A9A9}
\title{CED: Consistent ensemble distillation for audio tagging}
\name{Heinrich Dinkel, Yongqing Wang, Zhiyong Yan, Junbo Zhang, Yujun Wang}
\address{Xiaomi Corporation, Beijing, China}
\begin{document}
%\ninept
%
\maketitle
\begin{abstract}
% This study explores using augmentation and knowledge distillation (KD) for improving audio classification and reducing model sizes on the Audioset benchmark.
Augmentation and knowledge distillation (KD) are well-established techniques employed in audio classification tasks, aimed at enhancing performance and reducing model sizes on the widely recognized Audioset (AS) benchmark. 
Although both techniques are effective individually, their combined use, called consistent teaching, hasn't been explored before. 
This paper proposes CED, a simple training framework that distils student models from large teacher ensembles with consistent teaching.
To achieve this, CED efficiently stores logits as well as the augmentation methods on disk, making it scalable to large-scale datasets.
Central to CED's efficacy is its label-free nature, meaning that only the stored logits are used for the optimization of a student model only requiring 0.3\% additional disk space for AS.
The study trains various transformer-based models, including a 10M parameter model achieving a 49.0 mean average precision (mAP) on AS. 
Pretrained models and code are available \href{https://github.com/RicherMans/CED}{online}.

% While conssitent teaching has been studied in the vision domain, there has been no such study in the field of audio tagging.
\end{abstract}
\section{Introduction}
\label{sec:intro}

Audio tagging (AT) is a task that categorizes sounds into a fixed set of event classes, e.g., a baby crying or water running.
Applications of AT systems include aid for the hearing impaired, general monitoring of sounds~\cite{xiong_construction,dinkel2023streaming} as well as additional targets for keyword spotting~\cite{dinkel22_interspeech,dinkel23_icassp}.
Enhancing performance and minimizing the size of AT systems is vital for practical deployment. 
We target performance and size enhancement through common methods: data augmentation and knowledge distillation (KD).

\begin{figure}[htbp]
    \centering
    \includegraphics[width=0.95\linewidth]{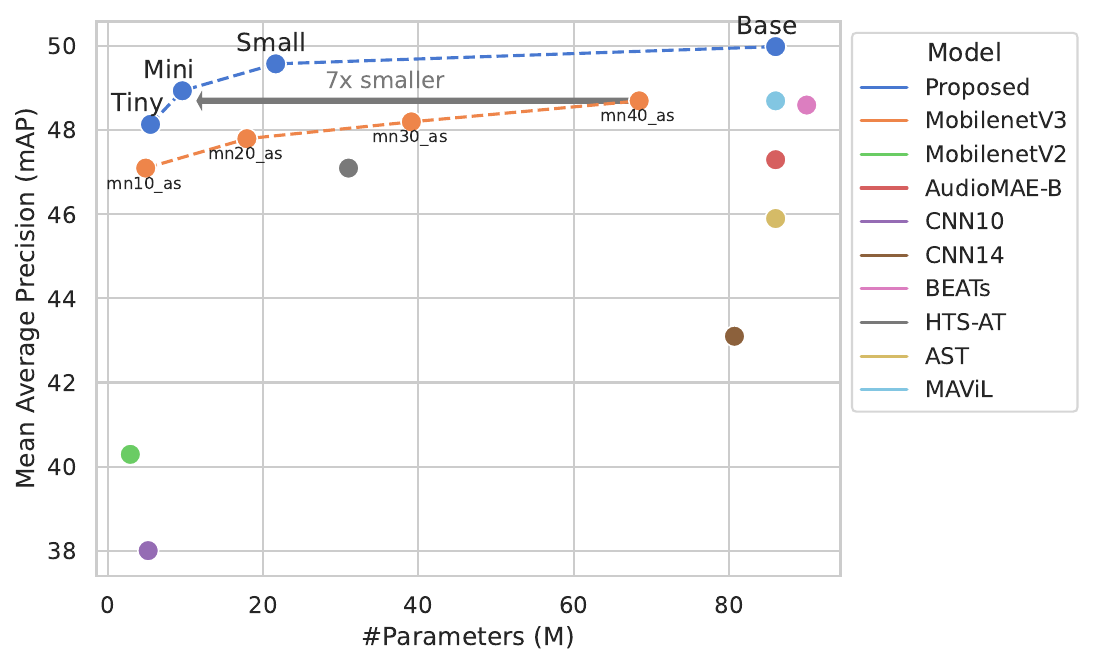}
     \caption{Our achieved performance in comparison to other works on Audioset (AS-2M). We reference results from the works in ~\cite{Chen2022beats,gong21b_interspeech,dinkel2022pseudo,chen2022hts,huang2022mavil,kong2020panns,huang2022masked}. } 
    \label{fig:enter-label}
\end{figure}
In KD, a large teacher model generates soft labels (logits) for a smaller student model to learn from. Typically, the objective of KD involves optimizing both the original hard labels and the logits together. 
Yet, recent research~\cite{dinkel2022pseudo} found that using only logits as training targets can significantly improve performance compared to the usual method.
By combining KD and data augmentation, also known as consistent teaching~\cite{beyer2022knowledge}, it has been suggested that performance can be further boosted. 
Surprisingly, no previous research has applied this approach to AT. 
We believe that the limited exploration of this approach is due to challenges in efficiently implementing KD.
Practically, there are two main ways to implement KD:
1. Online KD infers each soft label during training by forwarding a sample jointly through the teacher as well as the student.
2. Offline KD stores augmented samples as well as the teacher's soft labels on disk and reads them during student training.
\begin{figure}[htbp]
    \centering
    \includegraphics[width=0.95\linewidth]{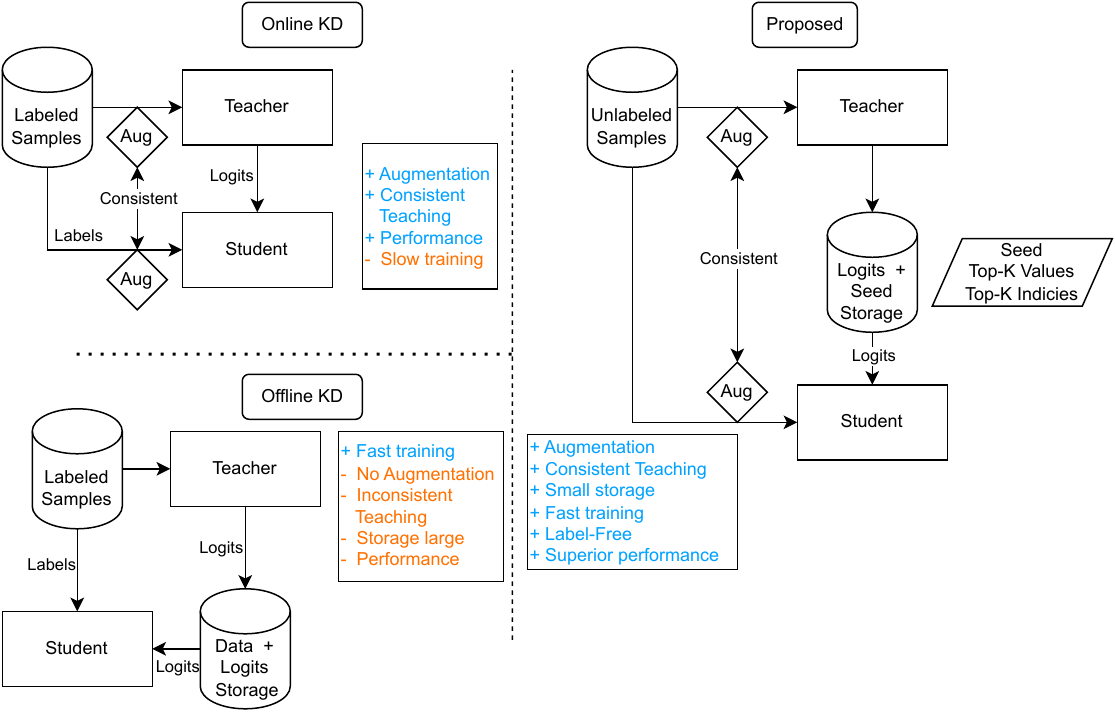}
    \caption{CED in comparison to standard on/off-line KD frameworks. 
    We first augment each sample using wave-level and spectrogram-level augmentations.
    Then we use an (ensemble) teacher model to predict scores for each respective sample and only store the seed which generated the augmentations and the top-k scores on disk.}
    \label{fig:intro_figure}
\end{figure}
Both these methods have pros and cons, detailed in \Cref{fig:intro_figure}.
Online KD is handicapped by its slow training speed because samples need to be sequentially forwarded through student and teacher, while offline KD can ``parallelize'' this process by first creating the teacher's data/logits. 
Conversely, offline KD struggles when handling substantial augmented data due to the storage demand of augmented samples. 
Thus, in practice, only logits from non-augmented training data are stored on disk.
Moreover, the performance of offline KD drops when (inconsistent) data augmentation techniques are used on the student's input, as demonstrated in previous studies~\cite{dinkel2022pseudo,schmid2023efficient}.
Our research shows (\Cref{ssec:consistent_teaching}) that consistent augmentation for both teacher and student inputs is crucial to improve performance.

We would like highlight key distinctions between our research two comparable works, Efficient-AT~\cite{schmid2023efficient} and PSL~\cite{dinkel2022pseudo} works.
Efficient-AT~\cite{schmid2023efficient} has utilized standard offline KD to achieve state-of-the-art (SOTA) performance on Audioset~\cite{gemmeke2017audio} (AS), but due to the nature of offline KD, could not apply consistent teaching.
Further, work in~\cite{dinkel2022pseudo} showed that label-free online KD is feasible for AT, yet could only use simple teacher models (MobileNetV2), since larger models significantly slow down training.
However, to the best of our knowledge, there has been no previous work that combined KD and augmentation with label-free online KD.
Finally, this work is closely related to the computer vision work TinyViT~\cite{wu2022tinyvit}, which applied a similar training pipeline for image classification. Our contributions are as follows:
\begin{enumerate*}[label=(\Roman*)]
    \item We propose CED, a simple framework to efficiently store and access logits as well as augmentation methods. CED only requires a few bytes of storage per sample, making it scalable to large datasets.
    \item We introduce consistent teaching to AT, which improves performance and reduces the performance gap between teacher and student models.
    \item We show that the features of CED models are also transferable to other audio classification tasks.
\end{enumerate*}

% While transformers have recently reigned surpreme in terms of 

\section{Method}
\label{sec:method}

A trivial solution to enable consistent teaching is to store an entire augmented dataset for each epoch on disk, use a teacher to predict logits and save those logits.
However, this solution is impractical for sizeable datasets, such as AS, due to its considerable storage demand (500 GB per epoch).
Further, audio can be augmented on wave and spectrogram levels, meaning that one would need to save augmented waveforms as well as their respective spectrograms, further increasing the storage requirement.

Instead of storing augmented samples on disk, CED only stores the seed that generated each respective augmented sample.
Specifically, given a single audio sample $\mathbf{x} \in \mathbb{R}^{T_{w}}$ with $T_w$ taps, we first apply wave-level augmentations $\hat{\mathbf{x}} = \text{wavaug}(\mathbf{x}, \phi)$ using a random seed $\phi$ on the sample.
Then, a Mel-spectrogram $\mathbf{X} \in \mathbb{R}^{F \times T}$ is extracted from $\hat{\mathbf{x}}$ and augmented using spectrogram-level augmentations: $\mathbf{\hat{X}} = \text{specaug}(\mathbf{X}, \phi)$.
Instead of storing $\hat{\mathbf{x}}$ and $\hat{\mathbf{X}}$, we efficiently store $\phi$ on disk for each sample.

Then we use a teacher model $\mathcal{T}$ to predict logits $\mathbf{y}_T \in [0,1]^{C}$ from the given sample $\hat{\mathbf{X}}$ for $C$ classes during each epoch $e = 1,\ldots,E$.
Directly storing $\mathbf{y}_T$ for a dataset with $N$ samples, $C$ classes and $E$ epochs requires $NCE$ disk space.
In the case of AS, $N=2\times10^{6}, C=527$, results in approximately $4$ GB (float 32) of storage per epoch.

In our study, we've chosen to preserve only the most prominent $K$ logits $\hat{\mathbf{y}}_{top-k}$ for each sample, along with the corresponding label indices $\hat{\mathbf{y}}_{top-idx}$.
Overall, we save $\{\hat{\mathbf{y}}_{top-k}, \hat{\mathbf{y}}_{top-idx}, \phi\}$ as 16-bit float, 16-bit integer and 32-bit integer, respectively.
This approach leads to a storage requirement of $((K \times (2 \times 2)) + 4)$ bytes per logit.
For our chosen $K = 20$, this results in approximately 84 bytes, or around 80 MB per AS-2M epoch. 
Notably, this is significantly lower compared to the na\"ive solution that would demand 4 GB.
As we only retain $K$ logits on disk, we assume a remaining probability of 0 for each sample. 
While we investigated alternative methods for handling remaining probabilities, these methods did not yield noticeable improvements.

\section{Experiments}
\label{sec:experiments}

\subsection{Dataset}

Our training and evaluation dataset is AS~\cite{gemmeke2017audio}, which mainly contains 10-second-long audio clips labelled with 527 different sound event classes.
We collected 1,904,746 training samples and 18,299 evaluation samples sampled at 16 kHz.
The training set is split into two subsets, the balanced (AS-20K) subset with 21,155 samples, and the entire full (AS-2M) subset with 1.9M samples.
Our experiments are first run on the AS-20K subset to ascertain our method's effect and then the findings are applied by training on AS-2M.

\subsection{Models}

\paragraph*{Student models}

The present study employs four vision transformer (ViT)-based architectures (see \Cref{tab:models}), namely (CED-) Tiny, Mini, Small, and Base~\cite{dosovitskiy2021an}, all of which closely follow ViT (pre-norm + GeLU activation).
Each model's setup follows~\cite{dinkel2023streaming}, where we employ 64-dimensional banks at a 16 kHz sampling rate, extracted within a 32 ms window and a 10 ms hop. 
These filterbanks are first normalized using batch normalization. 
From these spectrograms, we extract non-overlapping patches with a size of $16 \times 16$, resulting in $252 = 62 \times 4$ patches for a 10 s input in time/frequency, respectively. 
We use absolute positional embeddings, where time and frequency are independently modeled to allow these models to handle variable-sized inputs during inference.
\begin{table}[htbp]
    \centering
    \begin{tabular}{l|rrrr}
        \toprule
        Model &  \# Parameter & Embed & MLP & \#Heads \\
        \midrule
        Tiny & 5.5 M & 192 & 768 & 3  \\
        Mini &  10 M &  256 & 1024 & 4 \\
        Small & 22 M & 384 & 1536 & 6 \\
        Base & 86 M & 768 & 3072 & 12 \\
        \bottomrule
    \end{tabular}
    \caption{The utilized models in this study. ``Embed'' refers to the embedding dimension, ``MLP'' to the dimension of each block's multi-layer perceptron and ``\#Heads'' stands for the number of independent attention mechanisms.}
    \label{tab:models}
\end{table}
In order to achieve competitive performance~\cite{schmid2023efficient}, we use the masked autoencoder paradigm (MAE) to pre-train all teacher and student models~\cite{huang2022masked,dinkel2023streaming} on AS.

\paragraph*{Teacher model}

Inspired by \cite{schmid2023efficient}, we use an ensemble of differently-sized transformer models to predict labels.
Specifically, the ensemble consists of two ViT-Base models and three ViT-Large models, which all have been independently trained on AS.
Since CED requires consistency between the input features, the Mel-spectrogram configuration of each teacher is identical to a student.
The ensemble teacher model achieves an mAP of 50.1 on the evaluation set of AS.

\subsection{Augmentations and Logits}
\label{ssec:augmentation}

Even though our pipeline can support a plethora of augmentation methods, we keep the augmentations in this work simple.
In the waveform domain, we use sample-level shifting of the signal.
Further, in the spectrogram domain, we use SpecAug~\cite{park19e_interspeech}, masking at most 192 time-frames and 24 frequency banks.
On AS-20K we additionally apply mixup~\cite{zhang2018mixup} with $\lambda = 0.5$.
In order to further conserve storage space, we only save a certain amount of epochs and cycle over the dataset during training, which we set to $E=40$ for AS-20K and $E=10$ for AS-2M.
Further, we use $K=20$ as the default for all experiments, leading to an overall size of 1.5 GB, or 0.3\% of the AS-2M data.
% We will provide an investigation of
A storage requirement comparison between CED and previous works can be seen in \Cref{tab:storage}.

\begin{table}[htbp]
    \centering
    \begin{tabular}{ll|rr}
    \toprule
         Method & Aug? & AS-20K & AS-2M \\
         \midrule
         Na\"ive & \xmark &  42 & 3800  \\
         Efficient-AT~\cite{schmid2023efficient} & \xmark & 21  & 1900 \\
         \hline
         Proposed ($K=20$) & \cmark & \textbf{1.8} & \textbf{155} \\
         \bottomrule
    \end{tabular}
    \caption{Logit storage requirement per epoch given in Megabytes (MB). ``Na\"ive'' refers to storing all $C=527$ logits using float32 precision and ``Efficient-AT'' uses float16 precision. ``Aug?'' indicates consistent augmentation support.}
    \label{tab:storage}
\end{table}

\subsection{Setup}
\label{ssec:setup}

The majority of works on AS-2M use a balanced sampling strategy~\cite{kong2020panns} due to its long-tailed label distribution, which has been shown to have a negative impact on other datasets~\cite{performance_icassp2023_google}.
In line with the label-free nature of our work, we \textit{do not use a balanced sampling strategy}, since our method has no access to hard labels during training and thus sample randomly.
We train with an 8-bit Adam optimizer~\cite{dettmers2022optimizers} using a cosine learning rate decay scheduler and a maximal learning rate of 0.001 for the Tiny/Mini models and 0.0003 for the Small/Base models. 
We warmup the learning rate for 5,000 and 62,500 batches for AS-20K and AS-2M, respectively and decay the learning rates to 10\% of their maximal value over the training period.
We use the standard binary cross entropy loss (BCE) between the student's predicted logits and the teacher's logits as the training objective and the main evaluation metric is the mean average precision (mAP).
Training runs for 300 epochs with a batch size of 32 on AS-20K and for 120 epochs with a batch size of 128 on AS-2M.
Overall, logit extraction takes 30 hours and training takes at most 4 days on a single A100 GPU, depending on the model size.
We use AS-20K for model analysis and ablation studies.
% On a Tesla V100, our entire training pipeline takes at most 10 days for ViT-B
The neural network back-end is implemented in Pytorch~\cite{PaszkePytorch} and the source code with pretrained checkpoints is publicly available\footnote{\url{https://github.com/RicherMans/ced}}.

\section{Results}
\label{sec:results}

\subsection{Consistent teaching}
\label{ssec:consistent_teaching}

% Many previous works~\cite{dinkel2022pseudo,schmid2023efficient} have shown that utilizing augmentation on top of teacher-student training worsens performance instead of improving it.
% The reason for this phenomenon is the inconsistency between the inputs of a teacher and their student: 
% augmentation is only applied to the student inputs and not to the teacher inputs.
Here, we provide evidence that using consistent augmentations between teacher and student is crucial to improving performance.
The results of our experiment can be seen in ~\Cref{tab:consistency_influence}, where $\mathcal{T}_{Aug}$ represents using augmentation during logit prediction and $\mathcal{S}_{Aug}$ represents applying augmentation (see ~\Cref{ssec:augmentation}) on the student's input.
% Our findings are aligned with previous works~\cite{schmid2023efficient}, which show that inconsistency between teacher and student inputs is a major hindrance to better performance.
In summary, introducing augmentation to student inputs produces minor performance changes (within -0.8 to +0.2 mAP) across models, in line with~\cite{schmid2023efficient}. 
Conversely, augmenting only teacher inputs yields noteworthy enhancements, surpassing the baseline by more than 2 mAP points.
Finally, when using CED, which applies consistent teacher-student augmentation, substantial performance improvements emerge. 
These gains range from 5 to 7 mAP points over the baseline.

\begin{table}[tb]
    \centering
    \begin{tabular}{ll|rrrr}
    \toprule
        $\mathcal{T}_{Aug}$ & $\mathcal{S}_{Aug}$ & Tiny & Mini & Small & Base \\
    \midrule
        \xmark & \xmark & 28.52 & 30.52 & 32.28 & 37.87 \\
        \xmark & \cmark & 28.77 & 30.35 & 32.54 & 37.09 \\
        \cmark & \xmark & 31.75 & 33.45 & 34.30 & 39.03 \\
        \hline
        \cmark & \cmark &  36.47 & 38.50 & 41.55 & 43.97 \\
        % Tiny & 28.52 & 28.77 & 35.60 \\
        % Mini & 30.52 & & 37.57 \\
        % Small & 32.28 & 32.54 & 40.91 \\ 
        % Base & 37.87 & 37.99 & 43.41  \\
    \bottomrule
    \end{tabular}
    \caption{Impact of consistent training between teacher and student. The training data is AS-20K and values represent mAP on the AS evaluation set, where higher is better.}
    \label{tab:consistency_influence}
\end{table}

\subsection{Impact of saving $K$ logits}

\begin{figure}[tb]
    \centering
    \includegraphics[width=0.95\linewidth]{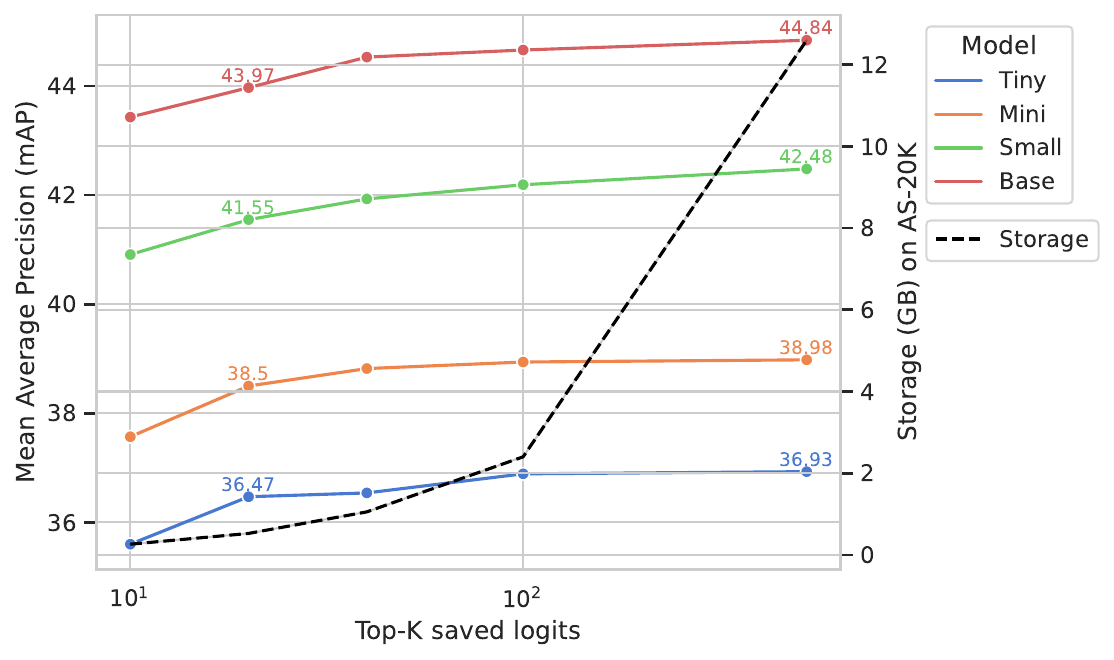}
    \caption{Performance and storage impact of different $K$ in AS-20K. We depict the achieved mAP for $K=20$ (proposed) and $K=527$ (best). Best viewed in color.}
    \label{fig:impact_of_k}
\end{figure}

\Cref{fig:impact_of_k} presents our findings on the AS-20K dataset using $K = \{10,20,40,100,527\}$, while setting $E=300$.
Generally, larger $K$ correlates with improved performance, though this improvement comes at the cost of heightened storage needs.
The performance gap between $K=10$ and $K=20$ (proposed) is significant, showing a 1-point mAP difference, while the storage impact remains minimal. 
Note that for $K=527$, the storage demand increases to 12 GB, an additional 190\% of the original dataset size of 6.3 GB.
In summary, the value of $K$ can be customized according to the dataset size and storage availability, given its direct influence on both performance and storage demands.

\subsection{Main results}

\begin{table}[tb]
    \centering

    \begin{tabular}{lll|rrr}
    \toprule
    & Model & \#Par (M) & AS-20K & AS-2M \\
    \midrule
   \parbox[t]{1mm}{\multirow{12}{*}{\rotatebox[origin=c]{90}{Baseline}}} & CNN14~\cite{kong2020panns} & 81 & 27.8 & 43.1 \\
    & MobileNetV2~\cite{dinkel2022pseudo} & 2.9 & 35.5 & 40.3 \\ 
    & HTS-AT~\cite{chen2022hts} & 31 & - & 47.1 \\ 
    & AST~\cite{gong21b_interspeech} & 86 & 34.7 & 45.9 \\ 
    & MaskSpec~\cite{chong2023masked} & 86 & 32.3 & 47.1 \\ 
    & BEATs~\cite{Chen2022beats} & 90 & 38.9 & 48.6\tablefootnote{On our evaluation split the model obtains 46.6.}\\
    & AudioMAE-B~\cite{huang2022masked} & 86 & 37.0 & 47.3 \\ 
    & ConvNeXt~\cite{pellegrini2023adapting} & 28 & - & 47.1 \\
    & MN10-AS~\cite{schmid2023efficient} & 4.9 & - & 47.1 \\ 
    & MN20-AS~\cite{schmid2023efficient} & 18 & - & 47.8 \\ 
    & MN40-AS~\cite{schmid2023efficient} & 68 & - & 48.7 \\ 
    % Color needs a switch here
    % {\color{xiaomi_gray}CAV-MAE~\cite{}} & {\color{xiaomi_gray}86} & {\color{xiaomi_gray}37.7} & {\color{xiaomi_gray}46.6} \\
    & {\color{xiaomi_gray}MAViL~\cite{huang2022mavil}} & {\color{xiaomi_gray}86} & {\color{xiaomi_gray}41.8} & {\color{xiaomi_gray}48.7} \\
    \hline
    \parbox[t]{1mm}{\multirow{4}{*}{\rotatebox[origin=c]{90}{CED}}} & Tiny & {5.5} & 36.5 & 48.1 \\
    % \quad + 30K &  & 39.9 &  \\
    & Mini & {10} &  38.5 & 49.0 \\
    % \quad + 30K & & & \\
    & Small &{22} & 41.6  & 49.6 \\
    % \quad + 30K & & 42.1 & \\
    & Base & {86} & \textbf{44.0} & \textbf{50.0} \\
    % \quad + 30K & & 45.7 &  \\
    \bottomrule
\end{tabular}
    \caption{Main results on AS-20K and AS-2M. 
    Models highlighted in grey have been trained with multi-modal supervision (Audio + Visual). Best in bold. 
    }
    \label{tab:main_results}
\end{table}

The main results of CED when compared to previous work can be seen in \Cref{tab:main_results}. 
Notably, our Mini model showcases superior performance, achieving an mAP of 49.0, while utilizing a mere 10 million parameters. 
Moreover, our single Base model with an mAP of 50 exhibits only a slight performance gap in comparison to the 5-way ensemble teacher with 50.1.

\vspace{-2mm}
\subsection{Transfer to downstream tasks}

Here we investigate whether CED-trained features are transferable to other downstream tasks, specifically for sound event detection (FSD50K, DCASE16) and acoustic scene classification (ESC-50).
To assess this, we employ the HEAR~\cite{turian2022hear} benchmark, which employs a linear classifier atop extracted features.
For all experiments, we extract features from the penultimate layer of our model by mean averaging all patches.
Results can be seen in~\Cref{tab:hear}, where CED-trained models are compared against alternative AS-based approaches.
CED-trained models can be seen to perform well across a variety of sound-related downstream tasks.

\begin{table}[tb]
    \centering
    \begin{tabular}{l|rrrrrrr}
    \toprule
        Model & FSD50K & ESC-50 & DCASE16\\
        \midrule
        CNN14 & - & 90.85 & 0.0 \\
        Eff-B2 & 60.71  &  93.45	 &  79.01 \\
        PaSST & 64.09 & 94.75 & 78.79 \\
        MN-40AS & 63.12 & 96.15 &  81.30 \\
        \hline
        Tiny & 62.73 & 95.80 & 88.02  \\
        Mini  & 63.88 & 95.35 & 90.66 \\
        Small & 64.33 & 95.95 &  91.63 \\
        Base & \textbf{65.48} & \textbf{96.65} & \textbf{92.19} \\
        \bottomrule
    \end{tabular}
    \caption{Linear evaluation results on the HEAR benchmark for sound detection tasks in comparison to previous works. Additional results are publicly \href{https://hearbenchmark.com/hear-leaderboard.html}{available}. Best in bold.}
    \label{tab:hear}
\end{table}

% \subsection{Impact of data sampling}

% Due to the imbalanced nature of labels in Audioset, most previous works~\cite{kqq_panns} are required to use a stratified sampling strategy.
% A less focused? aspect of balanced sampling is its impact on the input/output (I/O) of a machine.
% Balanced sampling can only be effectively utilized on disks with fast random access, such as solid-state drives (SSDs).

% In contrast, since our work has no access to hard labels, we cannot sample according to the number of samples present per class.

% \section{Limitations}
% \label{sec:limitations}

% CED requires to have consistent augmented views between teacher and student, and thus needs identical spectrogram front-ends between teacher and student.
% This, however, might not be optimal for some cases where one would like to improve the inference speed by i.e., using a larger hop size to decrease the number of time frames for the student model~\cite{liuxubo_simpf}.
% Further, our study focuses on data-augmentation methods, but our work can also be extended to model-specific augmentation methods, such as dropout~\cite{srivastava14a} and patch-out~\cite{koutini22passt}.

\vspace{-3mm}
\section{Conclusion}
\label{sec:conclusion}

This work introduced CED, a simple training framework for distilling AT models with consistent teaching.
Our work aims to efficiently distil a single model from an ensemble of large teacher models by storing the teacher model's logits as well as their respective augmentation method on disk.
Our results show that with CED, we can efficiently distil single models that are capable of achieving performance similar to large ensembles.
The Mini network can achieve an mAP of 49.0, outperforming previous studies by a significant margin, with a fraction of the number of parameters.
While this work focuses on transformer-based teacher and student models, it is important to note that CED is a general framework and can be used to distil other network types.

\small
% -------------------------------------------------------------------------
\bibliographystyle{IEEEbib}
\bibliography{refs}

\end{document}